# The spectrum of critical exponents in $(\vec{\phi}^2)^2$–theory in $d = 4 - \epsilon$ dimensions — Resolution of degeneracies and hierarchical structures

Stefan K. Kehrein[*]

Institut für Theoretische Physik,
Ruprecht–Karls–Universität,
D-69120 Heidelberg, Germany

June 18, 1995

**Abstract**

The spectrum of critical exponents of the $N$–vector model in $4 - \epsilon$ dimensions is investigated to the second order in $\epsilon$. A generic class of one–loop degeneracies that has been reported in a previous work is lifted in two–loop order. One– and two–loop results lead to the conjecture that the spectrum possesses a remarkable hierarchical structure: The naive sum of any two anomalous dimensions generates a limit point in the spectrum, an anomalous dimension plus a limit point generates a limit point of limit points and so on. An infinite hierarchy of such limit points can be observed in the spectrum.

[*]E–mail: kehrein@marvin.tphys.uni-heidelberg.de

# 1 Introduction

Relatively little is known about properties of the spectrum of critical exponents in $d > 2$ dimensional field theories. Conformal symmetry yields insufficient constraints for exact solutions in $d > 2$ dimensions and one has to rely on perturbation expansions. In two recent publications [6, 7] we have therefore studied the spectrum of critical exponents of the full operator algebra of composite operators in the critical $N$–vector model in $d = 4 - \epsilon$ dimensions in the $\epsilon$–expansion. In one–loop order, a solution of the spectrum problem in terms of an operator $V$ of anomalous dimensions could be given. The diagonalization of $V$ is a standard exercise, though it is still nontrivial to make general statements about the spectrum [7]. For an alternative approach and further information about the spectrum see also ref. [3].

Somehow surprising was the result derived from $V$ that there exists an infinite degeneracy in the spectrum of non–derivative composite operators. That is an infinite number of non–derivative eigenoperators (quasiprimary operators[1]) have vanishing anomalous dimensions in order $\epsilon$. In fact, in a sense made precise in ref. [7], for a given number of elementary fields in the composite operator and an increasing number of gradients, the one–loop spectrum of anomalous dimensions is dominated by this eigenvalue zero. This raised the question about the nature of these degeneracies.

In order to address this question in this paper, the one–loop calculation is extended to two–loop order for a particularly interesting subalgebra $\mathcal{C}_{sym}$ of the full operator algebra. Again one can derive a spectrum–generating operator that encodes all the information about two–loop eigenoperators and anomalous dimensions in $\mathcal{C}_{sym}$. The generic case will turn out to be the resolution of the one–loop degeneracies in two–loop order.

In combination with the previous results in one–loop order, this leads to the observation of a remarkable structure of limit points in the spectrum: One notices that the sum of any two eigenvalues in the spectrum generates a limit point in the spectrum. In this manner a naive sum rule for anomalous dimensions is recovered that generates limit points. Though this is essentially a conjecture, it can be proven explicitly for a small number of fields and is numerically very well supported in the general case. An interesting hierarchical structure of the spectrum follows naturally from this conjecture consisting of limit points of limit points of ... (and so on).

The structure of this paper is as follows. In section 2 we briefly sum up the one–loop results from our previous papers [6, 7] that are relevant for the purpose of this work. The calculation in extended to two–loop order in section 3. Using our notion of a spectrum–generating operator, it is straightforward to obtain eigenoperators and critical exponents in two–loop order. This is worked out in section 4 and we will see how the one–loop degeneracies are resolved. In addition, explicit evaluation of some critical exponents in the $\epsilon$–expansion provides a valuable consistency check by comparison with $1/N$–expansions by Lang and Rühl [12]. Since $(\vec{\phi}^2)^2$–theory describes the disordered phase and the nonlinear $\sigma$–model the ordered phase of the same second order phase transition, the critical exponents of all observables should be consistent in both expansion schemes. The conjecture regarding the hierarchical limit point structure of the spectrum is discussed in section 5. Section 6 is concerned with the structure of the two–loop eigenoperators explaining some new features as compared to one–loop order. The conclusions are summed up in section 7.

---

[1] We use the term "quasiprimary" in analogy to 2d CFTs for non–derivative eigenoperators, that is these operators are not the total spatial derivative of another operator. Clearly it is not possible to extend the concept of primary operators in a similar manner for $d > 2$.

## 2  Results in one–loop order

Let us briefly sum up some results of our previous papers [6, 7]. This will also define our notation.

The $N$–vector model in $d = 4 - \epsilon$ dimensions is defined by the Lagrangian

$$\mathcal{L} = \frac{1}{2}\,(\partial \vec{\phi})^2 + \frac{1}{2}\,m_0^2\,\vec{\phi}^2 + \frac{g_0}{4!}\,(\vec{\phi}^2)^2 \tag{1}$$

with the $N$–component field $\vec{\phi} = (\phi_1, \ldots, \phi_N)$. In this work we restrict ourselves to the case $N > 1$ and we will only be interested in a subalgebra $\mathcal{C}_{sym}$ of composite operators introduced in ref. [7]. This subalgebra allowed for the most complete classification in one–loop order [7] and will also lead to a manageable though still rather lengthy two–loop calculation. We expect that it still shows the generic features of the field structure of the $N$–vector model.

Composite operators in $\mathcal{C}_{sym}$ transform according to the following Young frames under internal SO($N$) and spatial SO(4) rotations

$$\begin{array}{ll} \text{SO}(N) & \boxed{1\,|\,2\,|\,\ldots\,|\,n} \\ \text{SO}(4) & \boxed{1\,|\,2\,|\,\ldots\,|\,l} \end{array} \tag{2}$$

where $n$ denotes the number of elementary fields and $l$ the number of gradients in the composite operator. $n$ and $l$ will generally have these meanings in the sequel. Operators in $\mathcal{C}_{sym}$ can be written as

$$O(\mathbf{t}, \mathbf{h}, \mathbf{c}) = t_{i_1 \ldots i_n}\, h_{\alpha_1 \ldots \alpha_l}\, O^{i_1 \ldots i_n}_{\alpha_1 \ldots \alpha_l}(\mathbf{c}) \tag{3}$$

where $t_{i_1 \ldots i_n}$ and $h_{\alpha_1 \ldots \alpha_l}$ are both completely symmetric and traceless tensors corresponding to SO($N$) and SO(4) structure, resp. Therefore $\mathcal{C}_{sym}$ contains no redundant operators with unphysical critical exponents (compare ref. [7]). The tensor operator $O^{i_1 \ldots i_n}_{\alpha_1 \ldots \alpha_l}(\mathbf{c})$ is a composite operator in the elementary fields $\phi_{i_1},\ldots,\phi_{i_n}$ with spatial derivatives $\partial_{\alpha_1},\ldots,\partial_{\alpha_l}$. It can generally be expressed like

$$O^{i_1\ldots i_n}_{\alpha_1\ldots\alpha_l}(\mathbf{c}) = \sum_{\{l_1,\ldots,l_n\,|\,l_j\geq 0,\sum l_j=l\}} \frac{c_{l_1\ldots l_n}}{l_1!\cdot\ldots\cdot l_n!}$$
$$\times \partial_{\alpha_1}\ldots\partial_{\alpha_{l_1}}\phi_{i_1}\,\partial_{\alpha_{l_1+1}}\ldots\partial_{\alpha_{l_1+l_2}}\phi_{i_2}\cdot\ldots\cdot\partial_{\alpha_{l-l_n+1}}\ldots\partial_{\alpha_l}\phi_{i_n} \qquad (4)$$

with coefficients $c_{l_1\ldots l_n}$. The factorials are introduced for later convenience. Due to the restriction to $\mathcal{C}_{sym}$ we need only consider completely symmetric coefficients $c_{l_1\ldots l_n}$. Therefore the tensor operators $O^{i_1\ldots i_n}_{\alpha_1\ldots\alpha_l}(\mathbf{c})$ can be mapped one to one on vectors in a Hilbert space

$$O^{i_1\ldots i_n}_{\alpha_1\ldots\alpha_l}(\mathbf{c}) \quad\longleftrightarrow\quad |\Psi(\mathbf{c})\rangle = \sum_{\{l_1,\ldots,l_n\,|\,l_j\geq 0,\sum l_j=l\}} c_{l_1\ldots l_n}\,a^\dagger_{l_1}a^\dagger_{l_2}\ldots a^\dagger_{l_n}|\Omega\rangle \qquad (5)$$

with creation operators $a^\dagger_j$ corresponding to multiplication with an elementary field with $j$ gradients acting on it

$$\Phi^{\alpha_1\ldots\alpha_j}_i \stackrel{\text{def}}{=} \frac{1}{j!}\partial_{\alpha_1}\ldots\partial_{\alpha_j}\phi_i \quad\longleftrightarrow\quad a^\dagger_j. \qquad (6)$$

The indices $i$ and $\alpha_1,\ldots,\alpha_j$ are omitted on the rhs since all indices are summed over with completely symmetric tensors in eq. (3) anyway. The corresponding annihilation operator $a_j$ is defined as the partial derivative with respect to the field $\Phi^{\alpha_1\ldots\alpha_j}_i$ and therefore $a_j$ and $a^\dagger_{j'}$ obey Bose commutation relations

$$[a_j,a^\dagger_{j'}] = \delta_{jj'}, \quad [a_j,a_{j'}] = [a^\dagger_j,a^\dagger_{j'}] = 0. \qquad (7)$$

$|\Omega\rangle$ in eq. (5) is a (formal) "zero fields" state.

The creation/annihilation operator notation is convenient since one can easily express the spectrum–generating operator $V^{(1lp)}$ in terms of them. In one–loop order we have shown in ref. [7]

$$V^{(1lp)} = \frac{1}{6}\sum_{l=0}^\infty \frac{1}{l+1}\sum_{j,k=0}^l a^\dagger_j a^\dagger_{l-j} a_k a_{l-k}. \qquad (8)$$

Eigenvectors $|\Psi\rangle$ of $V^{(1lp)}$ with eigenvalue $\lambda$ correspond via eqs. (5) and (3) to one–loop eigenoperators $O(\mathbf{t},\mathbf{h},\mathbf{c})$, $t_{i_1\ldots i_n}$ and $h_{\alpha_1\ldots\alpha_l}$ arbitrary completely symmetric and traceless tensors, with critical exponent

$$x = l + n\left(1 - \frac{\epsilon}{2}\right) + \tilde{g}_c\cdot\lambda + O(\epsilon^2). \qquad (9)$$

Here $\tilde{g}$ denotes a rescaled dimensionless coupling constant [23]

$$\tilde{g} = N_d\,\mu^{-\epsilon}\,g_0$$
$$N_d = \frac{2}{(4\pi)^{d/2}\Gamma(d/2)}, \qquad (10)$$

$\mu$ the usual mass scale of dimensional regularization. Up to order $\epsilon$ the critical coupling is

$$\tilde{g}_c = \epsilon \frac{6}{N+8} + O(\epsilon^2). \tag{11}$$

$V^{(1lp)}$ encodes the complete information about the one–loop spectrum of critical exponents and eigenoperators in $\mathcal{C}_{sym}$ in a compact form.

Properties of this spectrum have been worked out in detail in ref. [7]. Let us just sum up some results that will be relevant for comparison with the two–loop results:

- All eigenvalues $\lambda$ in eq. (9) are real and nonnegative, that is one–loop anomalous dimensions make composite operators "more" irrelevant.

- The symmetry group of $V^{(1lp)}$ is SO(2,1) left over from the full conformal symmetry group in four dimensions SO(5,1). Its generators are

$$\begin{aligned} D &= \sum_{j=0}^{\infty} (j+1)\, a_{j+1}^\dagger a_j \\ D^\dagger &= \sum_{j=0}^{\infty} (j+1)\, a_j^\dagger a_{j+1} \\ S &= \sum_{j=0}^{\infty} (j+1/2)\, a_j^\dagger a_j \end{aligned} \tag{12}$$

  corresponding to spatial derivatives, special conformal transformations and dilatations, resp.

- The spectrum consists of conformal blocks each generated by a conformally invariant eigenoperator (quasiprimary operator) annihilated by $D^\dagger$

$$D^\dagger |\Psi\rangle = 0. \tag{13}$$

  The number $d_{n,l}$ of conformally invariant operators with $n$ fields and $l$ gradients can be expressed in terms of a generating function

$$\sum_{l=0}^{\infty} d_{n,l}\, x^l = \frac{1}{\prod_{i=2}^{n}(1-x^i)} \tag{14}$$

  and increases asymptotically like $l^{n-2}/(n!(n-2)!)$ for fixed $n$.

- The number of eigenvalues zero for conformally invariant operators with $n$ fields and $l$ gradients is given by $d_{n,l-n(n-1)}$. For large $l$ "almost" all eigenvalues are zero for any fixed number of fields.

We attempt to find out whether these features are generic or an artefact of the relatively simple one–loop approximation by extending the calculation to two–loop order in the next section.

# 3 The operator for critical exponents in two–loop order

## 3.1 Diagrams and renormalization matrix

In order to renormalize a general composite operator $O$ in two–loop order, the one–particle irreducible diagrams in fig. 1 have to be calculated with the operator insertion $O$. In diagram 1 one has to extract the pole term $1/\epsilon$ and the finite term $O(\epsilon^0)$, in the order $g^2$ diagrams 2 to 8 one needs the double pole terms $1/\epsilon^2$ and the pole terms $1/\epsilon$.

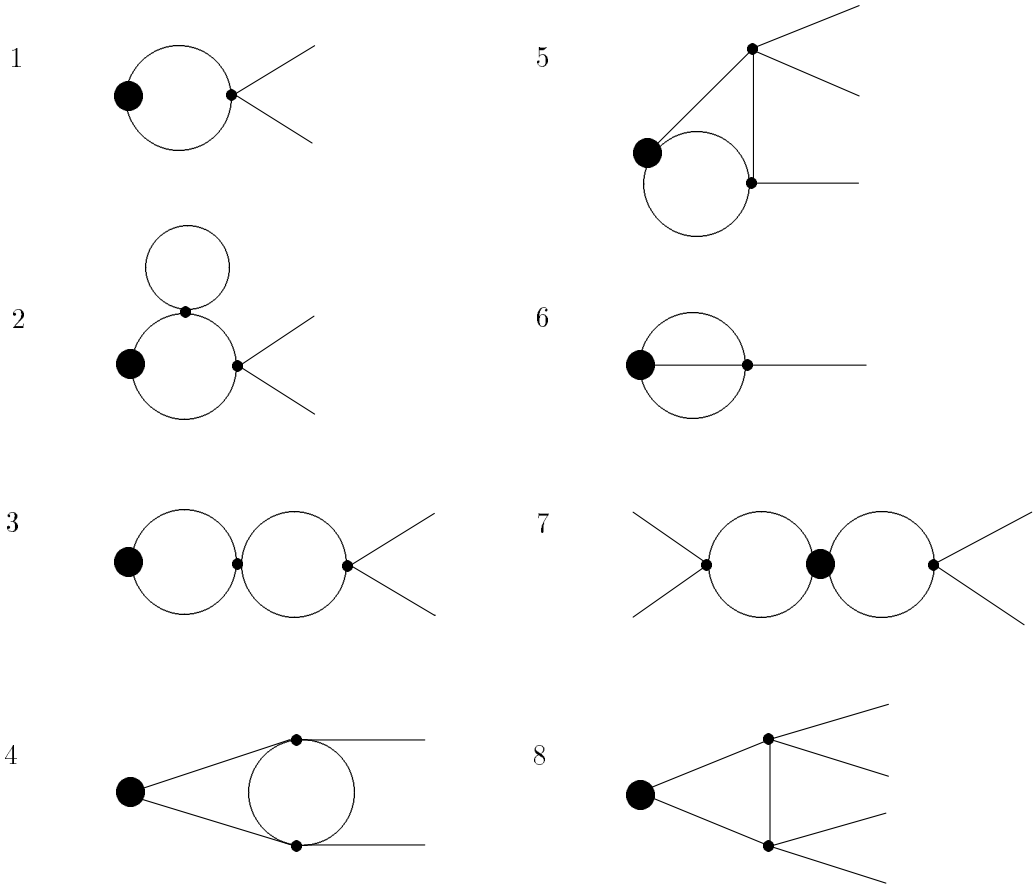

Figure 1: One–loop and two–loop diagrams 1 to 8. The bold point represents the operator insertion.

We will be working in the massless theory using dimensional regularization, therefore diagram 2 vanishes immediately. Another important simplification arises due to the restriction to operators in $\mathcal{C}_{sym}$. Diagrams 6 and 8, which could in principle lead to operator mixing with a different number of fields, do both not contribute in this case. For diagram 6 this is trivial since it vanishes identically because of the $O(N)$–tensor structure in eq. (3) and the condition $N > 1$. Diagram 8 cannot yield pole terms, i.e. it is finite: A divergent contribution can only occur if internal momenta generated by gradients in the composite operator cancel one of the propagators in the denominator. Since operators in $\mathcal{C}_{sym}$ are traceless in the spatial indices, this cannot happen here.

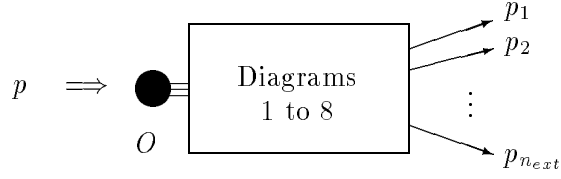

Figure 2: Composite operator insertion in a vertex function with $n_{ext}$ external legs.

We are left with the calculation of diagrams 1,3,4,5 and 7 for a general operator insertion from $\mathcal{C}_{sym}$. The details of this lengthy calculation can be found in the appendix. The main problem is to find expressions for the coefficients multiplying the pole terms which exhibit the dependence on the operator insertion in a simple manner.

The pole terms determine the renormalization factors of the composite operators in a standard way. Renormalized and bare vertex functions are connected via the renormalization matrix $Z_{A_k A_j}$

$$\Gamma^R_{A_k}(p_1,\ldots,p_{n_{ext}};p)\, Z_{A_k A_j}\, Z_\phi^{-n_{ext}/2} = \Gamma^B_{A_j}(p_1,\ldots,p_{n_{ext}};p). \tag{15}$$

Here $\Gamma^R_{A_k}$ denotes the renormalized vertex function with operator insertion $A_k$, similarly $\Gamma^B_{A_j}$ for the bare vertex function. $n_{ext}$ is the number of external legs, $p_1,\ldots,p_{n_{ext}}$ are the associated external momenta and $p$ is the momentum of the operator insertion, compare fig. 2. $Z_\phi$ in eq. (15) is the wave–function renormalization factor [23]

$$Z_\phi = 1 - \frac{\tilde{g}^2}{\epsilon}\frac{N+2}{144} + O(\tilde{g}^3). \tag{16}$$

The renormalization factor in eq. (15) expressed in terms of creation and annihilation operators follows from the Feynman diagrams calculated in appendix A

$$\begin{aligned} Z Z_\phi^{-n_{ext}/2} &= 1 + \tilde{g}\, X + \tilde{g}^2\, Y + O(\tilde{g}^3) \\ \Rightarrow\quad Z &= 1 + \tilde{g}\, X + \tilde{g}^2\, Y - \frac{n_{ext}}{2}\frac{\tilde{g}^2}{\epsilon}\frac{N+2}{144} + O(\tilde{g}^3) \end{aligned} \tag{17}$$

with

$$\begin{aligned} X &= -\frac{1}{6}\sum_{l=0}^\infty \sum_{k,j=0}^l \frac{1}{l+1}\, a^\dagger_j\, a^\dagger_{l-j}\, a_k\, a_{l-k} \\ &\quad \times\Big(\frac{1}{\epsilon} + \frac{1}{2}\ln\frac{\mu^2}{(p_j+p_{l-j})^2} - \frac{1}{2} + (S_{l+1} - \frac{1}{2}S_k - \frac{1}{2}S_{l-k}) + O(\epsilon)\Big) \end{aligned} \tag{18}$$

from eq. (78). $S_j$ is the sum

$$S_j = \sum_{k=1}^{j} \frac{1}{k} \ . \tag{19}$$

The lengthy expression for $Y$ is the sum of the contributions in order $\tilde{g}^2$ and can be found in the appendix, eq. (79). $Z_{A_k A_j}$ is now equivalent to the matrix element $<A_k|Z|A_j>$ in this creation/annihilation language.

In eq. (18) $p_j$ denotes the momentum of the field generated by $a_j^\dagger$. Similar logarithms of momenta can also be found in the expression for $Y$. However, one can get rid of these logarithms by a suitable subtraction scheme. Instead of the renormalization factor in eq. (17) we use the equivalent factor

$$\begin{aligned} Z &= (1 - \tilde{g}\, X_{fin}) \left( 1 + \tilde{g}\, X + \tilde{g}^2\, Y - \frac{n_{ext}}{2} \frac{\tilde{g}^2}{\epsilon} \frac{N+2}{144} \right) + O(\tilde{g}^3) \\ &= 1 + \tilde{g}\, X_{pole} + \tilde{g}^2\, (Y - X_{fin}\, X_{pole}) - \frac{n_{ext}}{2} \frac{\tilde{g}^2}{\epsilon} \frac{N+2}{144} + O(\tilde{g}^3). \end{aligned} \tag{20}$$

Here $X_{fin}$ is the finite part of $X$ from eq. (18) and $X_{pole}$ the pole term

$$X_{pole} = -\frac{1}{6\epsilon} \sum_{l=0}^{\infty} \frac{1}{l+1} \sum_{k,j=0}^{l} a_j^\dagger\, a_{l-j}^\dagger\, a_k\, a_{l-k} \tag{21}$$

$$X_{fin} = X - X_{pole}. \tag{22}$$

One can check in a straightforward calculation that all renormalization scale dependent logarithms in the combination $Y - X_{fin} X_{pole}$ cancel.

## 3.2 The two–loop mixing matrix in $\mathcal{C}_{sym}$

The mixing matrix (more precisely: mixing operator) follows in the usual manner from the renormalization matrix

$$M = \left( N_d\, \beta(\tilde{g})\, \frac{d}{d\tilde{g}}\, Z \right) Z^{-1}. \tag{23}$$

With the $\beta$–function [23]

$$N_d\, \beta(\tilde{g}) = -\epsilon\, \tilde{g} + \frac{N+8}{6}\, \tilde{g}^2 + O(\tilde{g}^3) \tag{24}$$

this is

$$\begin{aligned} M^{(2lp)} &= \frac{n}{2}\, \tilde{g}^2\, \frac{N+2}{72} - \epsilon\, \tilde{g}\, X_{pole} + \epsilon\, \tilde{g}^2\, X_{pole}^2 + \tilde{g}^2\, \frac{N+8}{6}\, X_{pole} \\ &\quad - 2\epsilon\, \tilde{g}^2\, (Y - X_{fin} X_{pole}) + O(\tilde{g}^3), \end{aligned} \tag{25}$$

where we have made use of the fact that in all vertex functions contributing here the numer of external fields is equal to the number of fields in the composite operator $n_{ext} = n$. All double pole terms in (25) cancel as they should do. By combining the appropriate terms one finally finds after some lengthy but simple algebra

$$M^{(2lp)} = \frac{n}{2}\eta + \tilde{g}_c V^{(1lp)} + \tilde{g}_c^2 \left(V_2^{(2lp)} + V_3^{(2lp)}\right) + O(\epsilon^3) \tag{26}$$

at the critical point [23]

$$\tilde{g}_c = \epsilon \frac{6}{N+8} + \epsilon^2 \frac{18(3N+14)}{(N+8)^3} + O(\epsilon^3). \tag{27}$$

$\eta$ follows from wave–function renormalization

$$\eta = \epsilon^2 \frac{N+2}{2(N+8)^2} + O(\epsilon^3). \tag{28}$$

In eq. (26) $V^{(1lp)}$ is the already known one–loop term eq. (8)

$$V^{(1lp)} = \frac{1}{6} \sum_{l=0}^{\infty} \frac{1}{l+1} \sum_{k,j=0}^{l} a_j^\dagger a_{l-j}^\dagger a_k a_{l-k}. \tag{29}$$

The two–loop terms can be split up in a two–particle interaction operator

$$\begin{aligned}V_2^{(2lp)} &= -\frac{N+6}{36} \sum_{l=0}^{\infty} \sum_{k,j=0}^{l} \frac{1}{l+1} \\ &\quad \times \left[1 - \frac{1}{2(l+1)} + \frac{1}{2}I(k<j)(S_k - S_j) + \frac{1}{2}I(k>j)(S_{l-k} - S_{l-j})\right] \\ &\quad \times a_j^\dagger a_{l-j}^\dagger a_k a_{l-k}\end{aligned} \tag{30}$$

with

$$I(\text{true}) = 1, \quad I(\text{false}) = 0 \tag{31}$$

and the three–particle interaction part given by

$$\begin{aligned}V_3^{(2lp)} &= -\frac{1}{9} \sum_{l=0}^{\infty} \sum_{k_1,j_1=0}^{l} \sum_{k_2=0}^{l-k_1} \sum_{j_2=0}^{l-j_1} \frac{1}{j_1+j_2+1} \frac{1}{k_1+k_2+1} \\ &\quad \times \bigg[\frac{1}{2}(S_{l-k_1-k_2} - S_{l+1}) \\ &\quad\quad + (S_{k_1+k_2+1} - \frac{1}{2}S_{k_1} - \frac{1}{2}S_{k_2})I(k_1+k_2+j_1+j_2 \geq l) \\ &\quad\quad + \frac{1}{2}(S_{l-j_1-j_2} - S_{l-k_1-k_2-j_1-j_2-1})I(k_1+k_2+j_1+j_2 < l)\bigg] \\ &\quad \times a_{j_1}^\dagger a_{j_2}^\dagger a_{l-j_1-j_2}^\dagger a_{k_1} a_{k_2} a_{l-k_1-k_2}.\end{aligned} \tag{32}$$

Notice that possible four–particle interaction terms have cancelled in eq. (25).

Eq. (26) is the key computational result of this paper. A right eigenvector $|\Psi(\mathbf{c})>$ of $M^{(2lp)}$ with $n$ fields and $l$ gradients

$$M^{(2lp)}|\Psi(\mathbf{c})> = \lambda |\Psi(\mathbf{c})> \tag{33}$$

corresponds to a two–loop eigenoperator $O(\mathbf{t},\mathbf{h},\mathbf{c})$ of the renormalization flow with anomalous dimension $\lambda$ as explained in sect. 2. Its critical exponent is

$$x = n\left(1 - \frac{\epsilon}{2}\right) + l + \lambda + O(\epsilon^3). \tag{34}$$

One can introduce an *operator of critical exponents* $X$ to code (34) into a compact expression

$$X = \left(1 - \frac{\epsilon}{2} + \frac{\eta}{2}\right)\hat{N} + \hat{L} + \tilde{g}\,V^{(1lp)} + \tilde{g}^2\left(V_2^{(2lp)} + V_3^{(2lp)}\right) \tag{35}$$

to be evaluated at the critical point $\tilde{g} = \tilde{g}_c$. Here $\hat{N}$ gives the number of fields

$$\hat{N} = \sum_{l=0}^{\infty} a_l^\dagger a_l \tag{36}$$

and $\hat{L}$ the number of gradients

$$\hat{L} = \sum_{l=0}^{\infty} l\, a_l^\dagger a_l. \tag{37}$$

## 4 Eigenoperators and critical exponents

### 4.1 General properties of the operator of critical exponents

The operator $X$ in eq. (35) encodes the complete two–loop spectrum of critical exponents and the corresponding eigenoperators in $\mathcal{C}_{sym}$. In this sense it gives the solution of the universal properties of the $N$–vector model in this subspace in two–loop order. Before we go on to obtain explicit information about the spectrum, let us establish some general properties of $X$.

First of all, $M^{(2lp)}$ commutes with the operators $S$ and the derivative operator $D$ from eq. (12). The commutator with $S$ is trivial and the commutator with $D$ follows from unbroken translation invariance (it can also be checked in a lengthy calculation). Hence

$$\begin{aligned}[] [S, X] &= 0 \\ [D, X] &= -D. \end{aligned} \tag{38}$$

In one–loop order the commutator $[D, V^{(1lp)}] = 0$ immediately implied the existence of another symmetry operator $D^\dagger$ since $V^{(1lp)}$ is hermitean. In two–loop order this simple argument no longer holds since $M^{(2lp)}$ is not hermitean: Both $V_2^{(2lp)}$ and $V_3^{(2lp)}$ are non–hermitean. Therefore the naive/canonical representation of the conformal group induced by $D, D^\dagger$ and $S$ is no longer a symmetry group on the operator algebra beyond one–loop order.

However, it can be shown [19] that it is possible to modify the canonical representation by terms in order $\epsilon$ yielding the explicit representation of the conformal symmetry group on $\mathcal{C}_{sym}$ in two–loop order. For the purpose of this paper it will be sufficient to argue that one only needs to find the non–derivative eigenoperators of $X$ that due to eq. (38) generate the whole spectrum.

Another property that cannot be extended beyond one–loop order is the lower bound zero for the anomalous dimensions [6] following from

$$V^{(1lp)} \geq 0. \tag{39}$$

E.g. operators without gradients with $n$ fields

$$t_{i_1...i_n} \phi_{i_1} \cdot ... \cdot \phi_{i_n}, \qquad t_{i_1...i_n} \text{ symmetric and traceless} \tag{40}$$

are easily seen to be eigenoperators of $X$ with critical exponent

$$\begin{aligned} x &= \left(1 - \frac{\epsilon}{2} + \epsilon^2 \frac{N+2}{4(N+8)^2}\right) n + \left(\frac{\epsilon}{N+8} - \epsilon^2 \frac{N^2 - 4N - 36}{2(N+8)^3}\right) n(n-1) \\ &\quad - \epsilon^2 \frac{2}{(N+8)^2} n(n-1)(n-2) + O(\epsilon^3) \end{aligned} \tag{41}$$

in agreement with former results [5]. For a fixed positive $\epsilon$ this exponent can become arbitrarily negative for a large number of fields since the term proportional to $n^3$ comes with a minus sign. However, this implies no danger for the stability of the nontrivial fixed point since the term proportional to $n^4$ in order $\epsilon^3$ is known to give a positive contribution [18]. It is after all not surprising that the asymptotic $\epsilon$–expansion generates terms with alternating signs in different orders.

In contrast for the $2 + \epsilon$–expansion in the $N$–vector model the situation is more worrisome. Both one–loop [20] and two–loop calculations [2] make a class of operators with many gradients more and more relevant for a given positive $\epsilon$. This stability problem for the nontrivial fixed point in the $2 + \epsilon$ expansion remains so far unresolved and seems to be generic for $2 + \epsilon$ expansions in all models [9, 10, 21, 14].

## 4.2 Results for two fields

For composite operators with $n = 2$ fields and $l$ gradients the critical exponents have already been calculated in the pioneering work of Wilson [22] up to order $\epsilon^2$. These results are rederived here in our formalism as a consistency check. In addition we also obtain the explicit structure of the eigenoperators that cannot be found in the calculational scheme of ref. [22].

*O(N) symmetric and traceless tensors:*

Non–derivative composite operators only exist for an even number of gradients $l$. Diagonalization of $X$ yields:

- $l = 0$ : Eigenoperator $t_{i_1 i_2} \phi_{i_1} \phi_{i_2}$, $t_{i_1 i_2}$ symmetric and traceless, with critical exponent

$$x = 2 - \epsilon \frac{N+6}{N+8} - \epsilon^2 \frac{N^2 - 18N - 88}{2(N+8)^3} + O(\epsilon^3). \tag{42}$$

- $l = 2, 4, \ldots$ : Eigenoperator $t_{i_1 i_2} h_{\alpha_1 \ldots \alpha_l} T_{i_1 i_2}^{\alpha_1 \ldots \alpha_l}$, $t_{i_1 i_2}$ and $h_{\alpha_1 \ldots \alpha_l}$ symmetric and traceless, where

$$T_{i_1 i_2}^{\alpha_1 \ldots \alpha_l} = \sum_{k=0}^{l} (-1)^k \binom{l}{k}^2 \partial_{\alpha_1} \ldots \partial_{\alpha_k} \phi_{i_1} \partial_{\alpha_{k+1}} \ldots \partial_{\alpha_l} \phi_{i_2}$$
$$-\tilde{g} \frac{N+6}{6} \frac{1}{l(l+1)} \partial_{\alpha_1} \ldots \partial_{\alpha_l} (\phi_{i_1} \phi_{i_2})$$
$$+\tilde{g} \sum_{\substack{p=2 \\ \text{even}}}^{l-2} c_{l,p} \, \partial_{\alpha_{p+1}} \ldots \partial_{\alpha_l} T_{i_1 i_2}^{\alpha_1 \ldots \alpha_p} + O(\tilde{g}^2). \tag{43}$$

with critical exponent

$$x = 2\left(1 - \frac{\epsilon}{2} + \frac{\eta}{2}\right) + l - \epsilon^2 \frac{N+6}{(N+8)^2} \frac{1}{l(l+1)} + O(\epsilon^3). \tag{44}$$

The coefficients $c_{l,p}$ in the eigenoperators with $l = 4, 6, \ldots$ are arbitrary. They result from the one–loop degeneracy of the anomalous dimensions for $l = 2, 4, \ldots$ and can only be fixed in a three–loop calculation.

The one–loop degeneracies in the spectrum are completely lifted by the terms in order $\epsilon^2$ in eq. (44) as already known from ref. [22]. A characteristic new feature is the spatial derivative of $\phi_{i_1} \phi_{i_2}$ in the eigenoperator in order $\tilde{g}$. Hence the eigenoperators are not invariant with respect to $D^\dagger$ from eq. (12)

$$D^\dagger T_{i_1 i_2}^{\alpha_1 \ldots \alpha_l}(0) \neq 0, \tag{45}$$

that is they are not *canonically* conformal invariant operators. We come back to this point in sect. 6.

*O(N) scalar tensors:*

Since diagram 6 does not contribute for $n = 2$, one can also deal with O(N) scalar tensors in this special case by using the appropriate combinatorial factors for the other diagrams. The operator of critical exponents on this subspace is

$$X\Big|_{n=2} = 2\left(1 - \frac{\epsilon}{2} + \frac{\eta}{2}\right) + l + \tilde{g}\frac{N+2}{12} \sum_{l=0}^{\infty} \frac{1}{l+1} \sum_{j,k=0}^{l} a_j^\dagger a_{l-j}^\dagger a_k a_{l-k}$$
$$-\tilde{g}^2 \frac{N+2}{12} \sum_{l=0}^{\infty} \sum_{k,j=0}^{l} \frac{1}{l+1} a_j^\dagger a_{l-j}^\dagger a_k a_{l-k} \tag{46}$$
$$\times \left[1 - \frac{1}{2(l+1)} + \frac{1}{2} I(k<j)(S_k - S_j) + \frac{1}{2} I(k>j)(S_{l-k} - S_{l-j})\right]$$

Like for the O(N) symmetric and traceless case one derives:

- $l = 0$ : Eigenoperator $\vec{\phi}^2$ with critical exponent

$$x = 2 - \epsilon \frac{6}{N+8} + \epsilon^2 \frac{(N+2)(13N+44)}{2(N+8)^3} + O(\epsilon^3). \tag{47}$$

- $l = 2, 4, \ldots$ : Eigenoperator $h_{\alpha_1 \ldots \alpha_l} \bar{T}^{\alpha_1 \ldots \alpha_l}$, $h_{\alpha_1 \ldots \alpha_l}$ symmetric and traceless, where

$$\begin{aligned}
\bar{T}^{\alpha_1 \ldots \alpha_l} &= \sum_{k=0}^{l} (-1)^k \binom{l}{k}^2 (\partial_{\alpha_1} \ldots \partial_{\alpha_k} \vec{\phi}) \cdot (\partial_{\alpha_{k+1}} \ldots \partial_{\alpha_l} \vec{\phi}) \\
&\quad - \tilde{g} \frac{1}{l(l+1)} \partial_{\alpha_1} \ldots \partial_{\alpha_l} \vec{\phi}^2 \\
&\quad + \tilde{g} \sum_{\substack{p=2 \\ \text{even}}}^{l-2} \bar{c}_{l,p} \, \partial_{\alpha_{p+1}} \ldots \partial_{\alpha_l} \bar{T}^{\alpha_1 \ldots \alpha_p} + O(\tilde{g}^2).
\end{aligned} \tag{48}$$

with critical exponent

$$x = 2\left(1 - \frac{\epsilon}{2} + \frac{\eta}{2}\right) + l - \epsilon^2 \frac{N+2}{2(N+8)^2} \frac{6}{l(l+1)} + O(\epsilon^3). \tag{49}$$

The same remarks as for the $O(N)$ symmetric and traceless tensors hold here too. In particular the critical exponents are consistent with ref. [22].

The special case with $l = 2$ gradients deserves some attention. In this case the critical exponent equals the dimension of space

$$x = d \tag{50}$$

and the corresponding scaling eigenoperator is proportional to the stress tensor

$$\Theta^{(m_1, m_2)} = h_{\mu\nu}^{(m_1, m_2)} \left[ \partial_\mu \vec{\phi} \cdot \partial_\nu \vec{\phi} - \left(\frac{1}{6} - \frac{\tilde{g}}{36}\right) \partial_\mu \partial_\nu \vec{\phi}^2 \right] + O(\tilde{g}^2) \tag{51}$$

as it should be. The magnetic quantum numbers $m_{1,2} = \pm 1/2$ label the symmetric and traceless tensors $h_{\mu\nu}^{(m_1, m_2)}$ and correspond to the two SO(3) sectors of the rotation group SO(4). The renormalized stress tensor follows via the renormalization matrix $Z$ and one can show after a short calculation

$$\begin{aligned}
\Theta^{R\,(m_1, m_2)} &= h_{\mu\nu}^{(m_1, m_2)} \left[ \partial_\mu \vec{\phi} \cdot \partial_\nu \vec{\phi} - \left(\frac{1}{6} - \frac{\tilde{g}}{36}\left(1 + \frac{\tilde{g}}{\epsilon} \frac{N+2}{12}\right)\right) \partial_\mu \partial_\nu \vec{\phi}^2 \right] \\
&\quad + O(\tilde{g}^2 \epsilon^0) + O(\tilde{g}^3).
\end{aligned} \tag{52}$$

This is consistent with the alternative definition of the renormalized stress tensor to be the response of the renormalized action in curved space with respect to a variation of the metric (for the two–loop renormalized action of scalar $\phi^4$–theory in curved space compare ref. [13]).

## 4.3 More complicated composite operators

It is not very illuminating to write down a complete list of eigenvalues of the operator of critical exponents for $n > 2$. All this information is contained in our operator $X$ anyway. We will only concentrate on some important features of the two–loop spectrum.

*Comparison with $1/N$–expansions:*

The nonlinear $\sigma$–model describes the second order phase transition in the universality class of the $N$–vector model coming from the ordered low–temperature phase. The $(\vec{\phi}^2)^2$–theory describes this transition coming from the high-temperature phase. Critical exponents of observables should therefore be consistent in the $1/N$–expansion in the nonlinear $\sigma$–model and the $4 - \epsilon$ expansion here. This idea has e.g. been successfully tested for the elementary field and the energy density in refs. [16, 17]. Since Lang and Rühl have derived a scheme to obtain critical exponents for composite operators in the $1/N$–expansion in the nonlinear $\sigma$–model for $2 < d < 4$ dimensions [12], it requires little extra work to check consistency also for more complicated composite operators. Besides this offers a valuable test for the correctness of the lengthy calculations in both approaches.

Tables 1 and 2 contain the critical exponents for the first non–derivative eigenoperators with $n = 3$ and $n = 4$ fields. We use the notation

$$x = n \left(1 - \frac{\epsilon}{2} + \frac{\eta}{2}\right) + l + \tilde{g}_c \, \alpha^{(1lp)} + \tilde{g}_c^2 \left(\frac{N+6}{36} \alpha_2^{(2lp)} + \frac{1}{9} \alpha_3^{(2lp)}\right) + O(\epsilon^3) \tag{53}$$

where the coefficients $\alpha^{(1lp)}, \alpha_2^{(2lp)}$ and $\alpha_3^{(2lp)}$ follow from $V^{(1lp)}, V_2^{(2lp)}$ and the three particle interaction $V_3^{(2lp)}$, resp.

| Gradients | $\alpha^{(1lp)}$ | $\alpha_2^{(2lp)}$ | $\alpha_3^{(2lp)}$ |
|:---:|:---:|:---:|:---:|
| $l = 0$ | 1 | $-3$ | $-3$ |
| $l = 2$ | $5/9$ | $-16/9$ | $-7/54$ |
| $l = 3$ | $1/6$ | $-13/12$ | $-5/96$ |
| $l = 4$ | $7/15$ | $-203/150$ | $61/1500$ |
| $l = 5$ | $2/9$ | $-391/360$ | $-91/1080$ |
| $l = 6$ | $3/7$ | $-5189/4410$ | $541/6860$ |
| | 0 | $-5/18$ | 0 |

Table 1: Non–derivative eigenoperators with $n = 3$ fields and $l \leq 6$ gradients. The critical exponents follows with eq. (53).

In the $1/N$–expansion Lang and Rühl obtained the critical exponents of composite operators with transformation properties (2) as continuous functions of the space dimension $d$ for $2 < d < 4$. One can expand these exponents to second order in $\epsilon$ for $d = 4 - \epsilon$ and compare the resulting coefficients with the coefficients in order $N^0$ and $1/N$ in eq. (53). We found perfect agreement of $\alpha^{(1lp)}$ and $\alpha_2^{(2lp)}$ for all composite operators contained in tables 1 and 2.[2] The coefficient $\alpha_3^{(2lp)}$ only contributes in order $1/N^2$ and cannot be compared therefore. The agreement of factors like $-\frac{5189}{4410}\frac{\epsilon^2}{N}$ gives good enough reason to expect consistency for general $l, n$.

| Gradients | $\alpha^{(1lp)}$ | $\alpha_2^{(2lp)}$ | $\alpha_3^{(2lp)}$ |
|---|---|---|---|
| $l = 0$ | 2 | $-6$ | $-12$ |
| $l = 2$ | $13/9$ | $-79/18$ | $-265/54$ |
| $l = 3$ | 1 | $-7/2$ | $-5/2$ |

Table 2: Non–derivative eigenoperators with $n = 4$ fields and $l \leq 3$ gradients.

*Resolution of degeneracies with eigenvalue zero:*

A surprising result in one–loop order are the degeneracies of eigenoperators with vanishing anomalous dimensions [7]. For two fields it was already well known that these degeneracies are lifted in order $\epsilon^2$ (compare sect. 4.2). In general we will find the same resolution for $n > 2$ fields.

Tables 3 and 4 collect the first critical exponents for non–derivative eigenoperators with $\alpha^{(1lp)} = 0$. One can easily establish

$$V^{(1lp)}|\Psi\rangle = 0 \quad \Rightarrow \quad \langle\Psi|V_3^{(2lp)}|\Psi\rangle = 0 \tag{54}$$

by using the structure of the eigenoperators with one–loop eigenvalue zero as found in closed form in ref. [7]. Hence $\alpha_3^{(2lp)} = 0$ and only $\alpha_2^{(2lp)}$ is tabulated.

Except for the case $n = 3$ with $l = 6$ or $l = 8$ gradients all one–loop degeneracies are resolved. In fact this resolution of degeneracies holds as far as we have tested it on a computer (up to at least $l = 22$ for arbitrary $n$).

*Resolution of degeneracies with nonzero eigenvalues:*

We observed another class of generic degeneracies for $n = 4$ fields and an odd number of gradients in ref. [7]. The complete solution for this part of the spectrum was recently given by Derkachov and Manashov in ref. [3]. In two–loop order we again notice from table 5 that the degeneracies are lifted. This also holds as far as we have tested it (up to at least $l = 22$).

---

[2] I thank K. Lang for making the explicit values of the $1/N$–coefficients based on the works cited in ref. [12] available to me.

| Gradients | $\alpha_2^{(2lp)}$ | |
|---|---|---|
| $l = 6$ | $-5/18$ | } still degenerate |
| $l = 8$ | $-5/18$ | |
| $l = 9$ | $-79/672$ | |
| $l = 10$ | $-23/84$ | |
| $l = 11$ | $-2599/21600$ | |
| $l = 12$ | $-0.268\ldots$ | |
| | $-0.0642\ldots$ | |

Table 3: The first non–derivative eigenoperators with $n = 3$ fields and vanishing one–loop eigenvalue $\alpha^{(1lp)} = 0$ ($\Rightarrow \alpha_3^{(2lp)} = 0$).

| Gradients | $\alpha_2^{(2lp)}$ |
|---|---|
| $l = 12$ | $-251/675$ |
| $l = 14$ | $-731/1944$ |
| $l = 15$ | $-5897/16200$ |
| $l = 16$ | $-0.379\ldots$ |
| | $-0.188\ldots$ |

Table 4: The first non–derivative eigenoperators with $n = 4$ fields and vanishing one–loop eigenvalue $\alpha^{(1lp)} = 0$ ($\Rightarrow \alpha_3^{(2lp)} = 0$).

| Gradients | $\alpha^{(1lp)}$ | $\alpha_2^{(2lp)}$ | $\alpha_3^{(2lp)}$ |
|---|---|---|---|
| $l = 3$ | $1$ | $-7/2$ | $-5/2$ |
| $l = 5$ | $1$ | $-199/60$ | $-161/60$ |
| $l = 7$ | $1$ | $-451/140$ | $-389/140$ |
| | $1/3$ | $-109/60$ | $3/20$ |
| $l = 9$ | $1$ | $-5317/1680$ | $-4763/1680$ |
| | $1/3$ | $-477/280$ | $11/105$ |
| | $5/9$ | $-8249/5040$ | $-1109/15120$ |

Table 5: Non–derivative eigenoperators with $n = 4$ fields and an odd number of gradients $l \leq 9$.

# 5 A conjecture about the structure of the spectrum

Let us put the information about the spectrum of anomalous dimensions together that we have obtained here and in refs. [6, 7]. Its graphical representation in fig. 3 leads to the following conjecture:

> Let $\lambda_1$ be the anomalous dimension of a non–derivative eigenoperator with $n_1$ fields, $\lambda_2$ likewise for an eigenoperator with $n_2$ fields. Then $\lambda = \lambda_1 + \lambda_2$ is a limit point in the spectrum of anomalous dimensions with $n_1 + n_2$ fields.

A similar conjecture has been formulated independently by Derkachov and Manashov in ref. [3].[3] We will later see how this gives rise to a remarkable infinite hierarchical structure of the spectrum.

The intuitive idea behind the conjecture follows an idea due to Parisi [1]. He argued that for composite operators like $\phi\,\partial_{\mu_1}\ldots\partial_{\mu_l}\phi$ in the limit of large spin $l \to \infty$, no further subtractions are necessary besides the wave function renormalization of the elementary fields $\phi$. Intuitively a large angular momentum "effectively" separates two space points. It seems to be possible to generalize this conjecture in the following manner: Let $O_1$ ($O_2$) be a non–derivative eigenoperator from $\mathcal{C}_{sym}$ with anomalous dimension $\lambda_1$ ($\lambda_2$) consisting of $n_1$ ($n_2$) fields and $l_1$ ($l_2$) gradients. Consider the composite operator made up of $O_1$ and $O_2$

$$P^l[O_1, O_2] = \sum_{k=0}^{l} m_k^l \, (D^k O_1)(D^{l-k} O_2) \tag{55}$$

with coefficients $m_k^l$ chosen as to make $P^l[O_1, O_2]$ canonically conformal invariant

$$m_k^l = (-1)^k \binom{l}{k} \frac{1}{(k - 1 + 2l_1 + n_1)!\,(l - k - 1 + 2l_2 + n_2)!} \;. \tag{56}$$

For finite $l$, $P^l[O_1, O_2]$ will generally not be an eigenoperator. But for $l \to \infty$ it approximates better and better a one–loop eigenoperator with an anomalous dimension $\lambda$ given by the naive sum of $\lambda_1$ and $\lambda_2$

$$\lambda = \lambda_1 + \lambda_2. \tag{57}$$

In this sense the derivatives $D$ in eq. (55) "divide" the two parts of the composite operator $P^l[O_1, O_2]$.

Though a closed proof of the above statement cannot be given, numerical calculations with eq. (55) strongly support these intuitive ideas for general $n$ in one–loop order. It is easy to check this and we will refrain from giving a lengthy list of explicit examples. One problem in the proof is the existence of the composite operators $P^l[O_1, O_2]$, since the sum in eq. (55)

---

[3] I thank the authors for informing me of their work prior to publication.

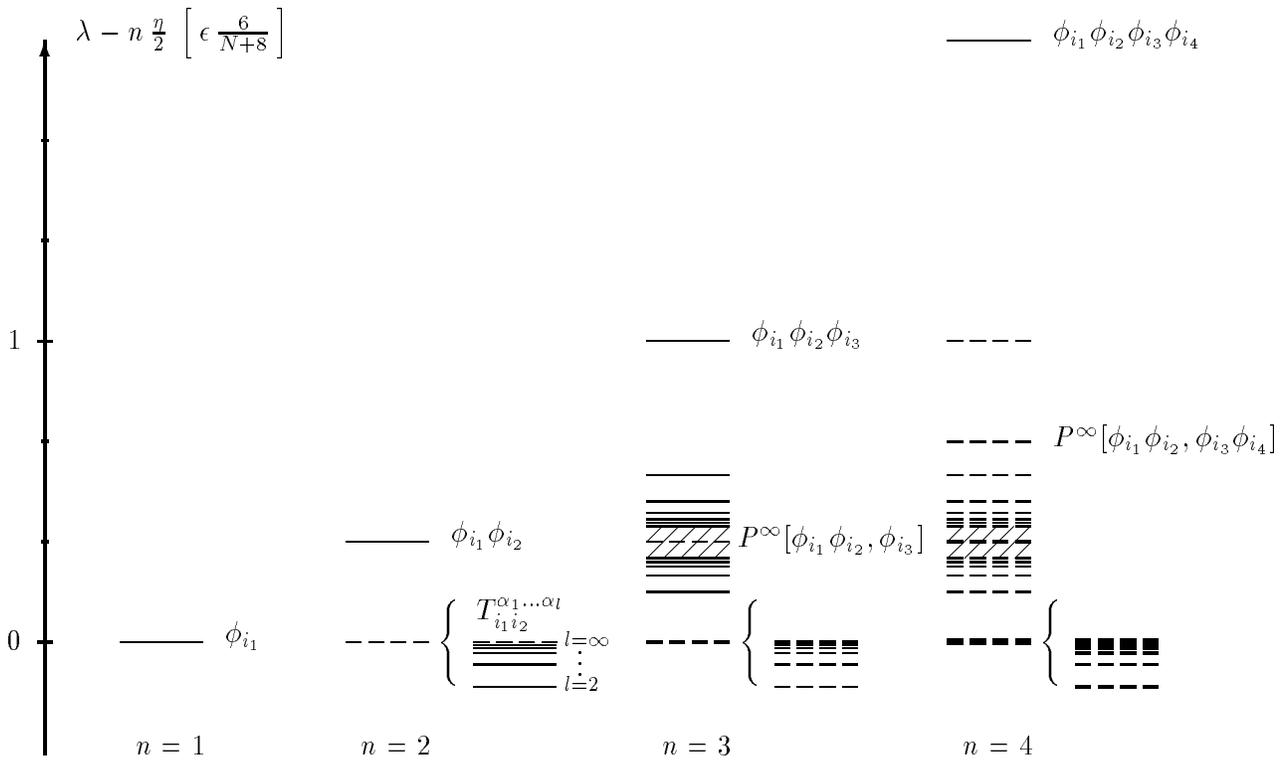

Figure 3: Anomalous dimensions $\lambda$ with wave function renormalization subtracted for non–derivative eigenoperators from $\mathcal{C}_{sym}$, eq. (2) with $n = 1$ to $n = 4$ fields. Full lines denote discrete eigenvalues, dashed lines are limit points in the spectrum. The linethickness of the dashed lines indicates the hierarchy of limit points. $O(\epsilon^2)$–splittings are qualitatively depicted with the brackets (only shown for the smallest eigenvalues). $n = 4$ also includes results from ref. [3]. For further explanations see text.

can vanish for particular values of $l$. In two–loop order things are more complicated since in general the coefficients $m_k^l$ are modified by terms in order $\epsilon$ making numerical calculations more involved. Still the results are in very good agreement with the conjecture.

It is interesting to notice that a similar structure of "limit points" can also be found in $1/N$–expansions: Since anomalous dimensions depend continuously on $d$ in $1/N$–expansions, it is more precise to speak of limit curves then. These have been observed for various classes of composite operators in ref. [11]. Hence one might expect that the conjecture holds to all orders in the $4 - \epsilon$ expansion.

Let us be more explicit by looking at composite operators with $n = 1$ to $n = 4$ fields from $\mathcal{C}_{sym}$, where the above ideas can in some cases be shown exactly. The anomalous dimensions $\lambda$ of these non–derivative operators are plotted in fig. 3. For convenience the contribution $n\,\eta/2$ coming from wave function renormalization has been subtracted every-where. In general only the order $\epsilon$ part of the anomalous dimension is depicted, except for the lowest–lying eigenvalues where a qualitative impression of the order $\epsilon^2$–splittings is given by the brackets. If one neglects these two–loop effects, fig. 3 is also the spectrum of anomalous dimensions in scalar $\phi^4$–theory in units of $[\epsilon]$.

- $n = 2$ : Going from left to right in the figure, the anomalous dimension of the elementary fields goes over in the family $T^{\alpha_1 \cdots \alpha_l}_{i_1 i_2}$ for $n = 2$. According to the conjecture $\lambda = 2\lambda_{\phi_i} = \eta$ appears as a limit point there indicated by the dashed line. It is approached by the discrete eigenvalues belonging to the eigenoperators $T^{\alpha_1 \cdots \alpha_l}_{i_1 i_2}$ from eq. (44). One other eigenvalue belonging to $\phi_{i_1} \phi_{i_2}$ appears.

- $n = 3$ : In agreement with the conjecture, there is the limit point $P^{\infty}[\phi_{i_1} \phi_{i_2}, \phi_{i_3}]$ for three fields. In fig. 3 this is the dashed line with $\lambda = \epsilon \frac{2}{N+8} + O(\epsilon^2)$. The discrete eigenvalues approaching it have already been worked out in ref. [6] in closed form. Close to the limit point the eigenvalues lie very densely and are indicated by the hatched areas.

   Furthermore each of the formerly discrete eigenvalues belonging to $T^{\alpha_1 \cdots \alpha_l}_{i_1 i_2}$ can be combined with an elementary field and becomes a limit point. The discrete eigenvalues approaching these limit points are not shown. In this manner a *limit point of limit points* is generated for $\lambda = 3\frac{\eta}{2}$ and depicted by the thicker dashed line.

- $n = 4$ : Only the largest of the discrete eigenvalues belonging to $\phi_{i_1} \phi_{i_2} \phi_{i_3} \phi_{i_4}$ is shown here. In addition it has already been observed in ref. [7] that the discrete eigenvalues for three fields appear here again with infinite multiplicities. Derkachov and Manashov have given a proof for this [3] and in addition shown that there is another limit point with $\lambda = \epsilon \frac{4}{N+8} + O(\epsilon^2)$. According to the conjecture this can be identified as $P^{\infty}[\phi_{i_1} \phi_{i_2}, \phi_{i_3} \phi_{i_4}]$ in fig. 3. The other limit points with $\lambda > O(\epsilon^2)$ obviously come from the combination of an elementary field with an eigenoperator with three fields.

   Order $\epsilon^2$–corrections resolve the infinite multiplicities and turn them into true limit points: If one does the two–loop calculation for sufficiently large $l$, one finds good evidence that eq. (57) holds in two–loop order too. Hence the former $P^{\infty}[\phi_{i_1}, \phi_{i_2} \phi_{i_3}]$ turns into a limit point of limit points here. Similarly for $\lambda = 2\eta$ one now finds a limit point of limit points of limit points.

It should be remarked that the above remarks have been proven in order $\epsilon$ in refs. [6, 7, 3]. In order $\epsilon^2$ and for $n = 3, 4$ the numerical evidence based on explicit diagonalizations is very good for non–degenerate subspaces. For subspaces that are degenerate in one–loop order, e.g. the lowest–lying eigenvalues, the approach to the limit points is slow (though steady) and an analytic proof would be desirable. In any case for a general number of fields some different technique will be necessary to prove the conjecture.

Summing up, one would propose that for an increasing number of fields $n$, higher and higher hierarchies of limit points are generated by the rules

$$\text{sum of two discrete eigenvalues} \longrightarrow \text{limit point}$$
$$\text{sum of a discrete eigenvalue and a limit point} \longrightarrow \text{limit point of limit points}$$

and so on. This produces an interesting infinite hierarchical structure in the spectrum of anomalous dimensions. For $n \to \infty$ the spectrum might even have fractal properties.

## 6 Conformal covariance in real space

### 6.1 Two–point correlation functions of composite operators

So far we have mainly been concerned with the spectrum of anomalous dimensions. In this section follows a short aside discussing the structure of the respective eigenoperators. The generic new feature in two–loop order are "anomalous" (i.e. order $\epsilon$) terms in the eigenoperators, compare for example eqs. (43) or (48). We want to demonstrate how this is connected with the idea of conformal symmetry, which is expected at the critical point of the $N$–vector model, thereby going beyond the well–established concept of global scale invariance. The most complete and rigorous treatment of this point is due to Schäfer [15].[4]

One important consequence of conformal invariance is that two non–derivative eigenoperators with different critical exponents are uncorrelated in a two–point function at the critical point

$$x_{A_1} \neq x_{A_2} \quad \Rightarrow \quad < A_1^R(r_1) A_2^R(r_2) >_c = 0. \tag{58}$$

Here $A_i^R$ are the renormalized eigenoperators. In principle the theorem (58) was already stated by Ferrara, Gatto and Grillo [4], but only in ref. [15] it became clear why the restriction to non–derivative eigenoperators is necessary. Of course all total spatial derivatives of $A_1$ and $A_2$ are also uncorrelated after $x_{A_1} \neq x_{A_2}$ has been shown for the non–derivative eigenoperators.

The terms in order $\epsilon$ in the two–loop eigenoperators will now turn out to be exactly the necessary correction terms to make eq. (58) hold to order $\epsilon$. In fact one can use eq. (58) as an alternative starting point to calculate these anomalous terms. It should be noted that such terms in the eigenoperators are no more "anomalous" than the anomalous dimension in the critical exponent. With respect to special conformal transformations they play a similar role like anomalous dimensions with respect to dilatations.

---

[4]In fact Schäfer establishes some conditions under which global scale invariance implies local scale invariance (i.e. conformal invariance). To the best of our knowledge there is no proof in the literature that all these conditions are actually fulfilled in the $N$–vector model, in particular his condition (Kii) regarding possible surface terms. Still we consider it plausible.

Let us make these points more explicit. We will calculate the two–point correlation function for composite operators up to terms in order $\epsilon$. For the purpose of this section it will prove to be sufficient to consider operators from $\mathcal{C}_{sym}$ that couple to maximum spin. One can achieve this by doing the calculation for composite operators $A_1, A_2$ with structure (68)

$$A_{1,2} = t^{(1,2)}_{i_1 \ldots i_{n_{1,2}}} \Delta_{\alpha_1} \ldots \Delta_{\alpha_{l_{1,2}}} A^{i_1 \ldots i_{n_{1,2}}}_{\alpha_1 \ldots \alpha_{l_{1,2}}}(\mathbf{c}^{(1,2)}), \tag{59}$$

$A^{i_1 \ldots i_{n_{1,2}}}_{\alpha_1 \ldots \alpha_{l_{1,2}}}(\mathbf{c}^{(1,2)})$ like in eq. (4) and $\Delta = (i/2, 0, 0, 1/2)$.

It is easy to see that up to order $g$, non–vanishing contributions in the two–point function (58) of O($N$)–traceless tensors can only arise for the same number $n$ of fields in both $A_1$ and $A_2$. Details of the calculation of the two–point correlation function for renormalized eigenoperators are worked out in appendix B. The result is

$$<A^R_1(r_1)\,A^R_2(r_2)>_c = \frac{2^{l_1+l_2+2n(1-\epsilon/2)}}{(4\pi)^{nd/2}} \frac{(\Delta \cdot r)^{l_1}(-\Delta \cdot r)^{l_2}}{(r^2)^{l_1+l_2+n(1-\epsilon/2)}}$$
$$\times \Bigg\{ \left(1 - \tilde{g}\left[\frac{\lambda_{A_1}}{\tilde{g}_c} + \frac{\lambda_{A_2}}{\tilde{g}_c}\right]\right)\left(\ln(\mu\,r) - \ln 2 - \frac{1}{2}\psi(2)\right)\mathcal{A}^{(0)}_0$$
$$+ \left(\epsilon + \tilde{g}\left[\frac{\lambda_{A_1}}{\tilde{g}_c} + \frac{\lambda_{A_2}}{\tilde{g}_c}\right]\right)\mathcal{A}^{(0)}_\epsilon - \frac{4}{3}\tilde{g}\,\mathcal{A}^{(1)}_\epsilon \Bigg\}$$
$$+ \tilde{g}\,O(\epsilon) + O(\tilde{g}^2) \tag{60}$$

with $r = r_2 - r_1$. $\lambda_{A_1}, \lambda_{A_2}$ are the anomalous dimensions of $A_1, A_2$, resp., and the amplitudes $\mathcal{A}^{(0)}_0, \mathcal{A}^{(0)}_\epsilon, \mathcal{A}^{(1)}_\epsilon$ can be found in appendix B. The correct scaling behaviour follows as usual by exponentiating logarithms.

## 6.2 Composite operators with two fields

*O(N) symmetric and traceless tensors:*
Consider the operator $S_l = t^{(1)}_{i_1 i_2} \Delta_{\alpha_1} \ldots \Delta_{\alpha_l} S^{\alpha_1 \ldots \alpha_l}_{i_1 i_2}$ for even $l$ with

$$S^{\alpha_1 \ldots \alpha_l}_{i_1 i_2} = \sum_{k=0}^{l}(-1)^k \binom{l}{k}^2 \partial_{\alpha_1}\ldots\partial_{\alpha_k}\phi_{i_1}\partial_{\alpha_{k+1}}\ldots\partial_{\alpha_l}\phi_{i_2} + \tilde{g}\,c_{l,0}\,\partial_{\alpha_1}\ldots\partial_{\alpha_l}(\phi_{i_1}\phi_{i_2}) \tag{61}$$

where $c_{l,0}$ is some arbitrary coefficient. $S_l$ is a one–loop eigenoperator for arbitrary $c_{l,0}$, but in two–loop order the correct scaling behaviour requires according to eq. (43)

$$c_{l,0} = -\frac{N+6}{6}\frac{1}{l(l+1)}. \tag{62}$$

Alternatively conformal covariance says

$$<\left[S_l\right]^R(0)\,[t^{(2)}_{i_1 i_2}\phi_{i_1}\phi_{i_2}]^R(r)>_c \stackrel{!}{=} 0. \tag{63}$$

From eq. (60) one finds after some calculation

$$\langle [S_l]^R(0) \, [t^{(2)}_{i_1 i_2} \phi_{i_1} \phi_{i_2}]^R(r) \rangle_c = \sum_{i_1, i_2} t^{(1)}_{i_1 i_2} t^{(2)}_{i_1 i_2} \times \left( \frac{2\epsilon}{l} - \tilde{g} \frac{2}{3l} + \tilde{g} \, c_{l,0} \, 2(l+1) \right)$$
$$+ \tilde{g} \, O(\epsilon) + O(\tilde{g}^2)$$
$$\stackrel{!}{=} 0, \tag{64}$$

therefore

$$\tilde{g} \, c_{l,0} = \frac{1}{l(l+1)} \left( \frac{\tilde{g}}{3} - \epsilon \right) \tag{65}$$

in agreement with eq. (62) at the critical point.

The anomalous terms in the two–loop eigenoperators can in this manner be derived from an order $g$ calculation: The coupling to the spatial derivative of $\phi_{i_1} \phi_{i_2}$ in the eigenoperator can be uniquely fixed by requiring that the eigenoperator is uncorrelated with $\phi_{i_1} \phi_{i_2}$, eq. (63). In a similar manner one could derive all the other anomalous terms $c_{l,p}$ in eq. (43) by requiring

$$\langle S_l^R(0) \, S_p^R(0) \rangle_c = 0. \tag{66}$$

*O(N) scalar tensors:*

For $O(N)$ scalar tensors with two fields we have also been able to find the scaling eigenoperators in sect. 4.2. Besides differing combinatorial factors, the calculation of the two–point function runs along the same lines as in appendix B. Again one can see explicitly that eq. (58) holds for all correlation functions of the operators $\bar{T}^{\alpha_1 \ldots \alpha_l}$ from eq. (48) with $\vec{\phi}^2$. In particular this is true for the correlation function of stress tensor (52) and energy density

$$\langle \Theta^{R \, (m_1, m_2)}(r_1) \, [\vec{\phi}^2]^R(r_2) \rangle_c = 0 + O(\epsilon^2). \tag{67}$$

This property holds only for the coefficient of the *improvement term* $\partial_\mu \partial_\nu \vec{\phi}^2$ in the stress tensor that we have found in sect. 4.2.

### 6.3 More complicated composite operators

So far we have looked at correlation functions of non–derivative operators where already the canonical critical exponents were different. A new feature arises for $n > 2$ fields since then one can have linearly independent non–derivative operators with the same number of fields and gradients. In this case the critical exponents in eq. (58) only differ by terms in order $\epsilon$ or higher.

Due to the generally very complicated structure of the two–loop eigenoperators for $n > 2$, we cannot give closed formulas here. However, we have checked explicitly that the two–point correlation functions of all the composite operators in tables 1 to 5 with each other are consistent with theorem (58).

# 7 Conclusions

The spectrum of anomalous dimensions in the operator subalgebra $\mathcal{C}_{sym}$ of the $N$–vector model has been investigated to order $\epsilon^2$ in the $d = 4 - \epsilon$ expansion. This analysis became possible using the concept of a spectrum–generating operator introduced in ref. [6]. The complete spectrum is encoded in the spectrum–generating operator and it is straightforward to extract anomalous dimensions and the corresponding eigenoperators.

An interesting consistency check in its own right is provided by comparison of critical exponents in the $\epsilon$–expansion here with results from $1/N$–expansions in the nonlinear $\sigma$–model for $2 < d < 4$ dimensions by Lang and Rühl [12]. Due to the resolution of one–loop degeneracies in two–loop order, we make the same observation as in the $1/N$–expansion that generically quasiprimary operators (non–derivative eigenoperators) generate independent conformal blocks. Another consistency check sensitive to the structure of the eigenoperators was provided by testing conformal covariance of two–point correlation functions of composite operators in sect. 6. Anomalous terms in the eigenoperators are found to play a similar role with respect to special conformal transformations as anomalous dimensions with respect to dilatations.

A key observation in this work is the infinite hierarchical structure in the spectrum as depicted in fig. 3. Such a structure follows naturally from the conjecture that for any two eigenoperators $O_1, O_2$, one can construct approximate eigenoperators $P^l[O_1, O_2]$ as in eq. (55) that become exact eigenoperators in the limit $l \to \infty$. The anomalous dimension is then just given by the naive sum of the two constituent anomalous dimensions. The intuitive idea behind this conjecture is that the two constituents $O_1, O_2$ become "separated" for large $l$. These ideas have been numerically and analytically well–supported, though a general proof is still an open problem. A similar conjecture was formulated independently by Derkachov and Manashov [3].

Thus the spectrum of this well–investigated $d$ dimensional field theory seems to possess a fascinating infinite hierarchical structure of limit points that allows for fractal properties in the limit of a large number of fields $n \to \infty$. It is an interesting question whether such properties of spectra could play a role in many–particle physics since this spectrum also has a simple interpretation as a local interaction problem on a certain two–dimensional manifold (compare the appendix of ref. [7]).

The author is indebted to F. Wegner for a great number of valuable and stimulating discussions. This provided many new insights on numerous occasions during the progress of this work. I would also like to thank K. Lang, S.E. Derkachov and A.N. Manashov for various valuable discussions and for making their results available to me prior to publication.

# Appendix A  Two–loop calculations

## A.1  Diagram 1

The one–loop diagram 1 in fig. 1 has to be evaluated for a general operator insertion $O$ from $\mathcal{C}_{sym}$. This one diagram is derived in more detail in order to demonstrate the typical steps of the calculation.

Due to SO(4) symmetry it is sufficient to do this calculation for the spatial tensor component with the smallest magnetic quantum numbers that can be written as (compare eq. (3))

$$O(\mathbf{t}, \mathbf{h}^{(-l/2,-l/2)}, \mathbf{c}) = t_{i_1 \ldots i_n} \, h^{(-l/2,-l/2)}_{\alpha_1 \ldots \alpha_l} \, O^{i_1 \ldots i_n}_{\alpha_1 \ldots \alpha_l}(\mathbf{c}) \tag{68}$$

with

$$h^{(-l/2,-l/2)}_{\alpha_1 \ldots \alpha_l} = \frac{1}{l!} \Delta_{\alpha_1} \cdot \ldots \cdot \Delta_{\alpha_l} \tag{69}$$

and the vector

$$\Delta = (i/2, 0, 0, 1/2). \tag{70}$$

$\Delta$ has the important property

$$\Delta^2 = 0 \tag{71}$$

that will simplify the calculation considerably.

For an operator insertion of this type the contribution of the one–loop diagram in fig. 4 follows by closing any two legs of the composite operator with the loop. Each such term leads to an integral

$$I_1(l_1, l_2) = 8 \times \left(-\frac{g_0}{4!}\right) \int \frac{d^d q}{(2\pi)^d} \frac{(\Delta \cdot q)^{l_1} (\Delta \cdot (p-q))^{l_2}}{q^2 (q-p)^2} \tag{72}$$

where $l_1$ and $l_2$ are the number of gradients ($\Delta \cdot \partial$) at the resp. legs of the composite operator. For traceless and symmetric O($N$)–tensors the combinatorial factor is 8 as included in eq. (72). This simple momentum integral can be evaluated by Laplace transformation and

$$\int \frac{d^d q}{(2\pi)^d} (\Delta \cdot q)^j e^{-\sigma q^2 + 2 p \cdot q} = \frac{1}{(4\pi)^{d/2}} \sigma^{-j-d/2} (\Delta \cdot p)^j e^{p^2/\sigma} \qquad (73)$$

where the property $\Delta^2 = 0$ has been used. One finds

$$I_1(l_1, l_2) = -\frac{g_0}{3 (4\pi)^{d/2} (p^2)^{\epsilon/2}} \Gamma(\frac{\epsilon}{2}) (\Delta \cdot p)^{l_1+l_2} B(l_1 + 1 - \frac{\epsilon}{2}, l_2 + 1 - \frac{\epsilon}{2}). \qquad (74)$$

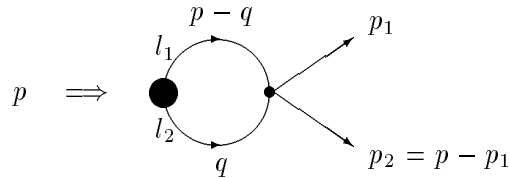

Figure 4: 1–Loop diagram with an operator insertion with total momentum $p$

Now we should use the renormalized coupling constant here instead of the bare coupling $g_0$. It is convenient to rescale as defined in eq. (10). The renormalized coupling constant is then $\tilde{g} Z_{\tilde{g}}$ with the renormalization factor [23]

$$Z_{\tilde{g}} = 1 + \frac{N+8}{6} \frac{\tilde{g}}{\epsilon} + O(\tilde{g}^2). \qquad (75)$$

Inserting everything in eq. (74) and expanding in $\epsilon$ yields

$$\begin{aligned}I_1(l_1, l_2) &= \left[ -\frac{\tilde{g}}{3} \left( \frac{1}{\epsilon} + \frac{1}{2} \ln \frac{\mu^2}{p^2} - \frac{1}{2} + (S_{l_1+l_2+1} - \frac{1}{2} S_{l_1} - \frac{1}{2} S_{l_2}) + O(\epsilon) \right) \right. \\ &\quad \left. - \frac{\tilde{g}^2}{3} \frac{N+8}{6} \left( \frac{1}{\epsilon^2} + \frac{1}{2\epsilon} \ln \frac{\mu^2}{p^2} - \frac{1}{2\epsilon} + \frac{1}{\epsilon} (S_{l_1+l_2+1} - \frac{1}{2} S_{l_1} - \frac{1}{2} S_{l_2}) + O(\epsilon^0) \right) \right] \\ &\quad \times \frac{l_1! l_2!}{(l_1 + l_2 + 1)!} (\Delta \cdot p)^{l_1+l_2} + O(\tilde{g}^3)\end{aligned} \qquad (76)$$

with the sums $S_j$ defined in eq. (19). Summed over all legs of the composite operator this gives the contribution of the one–loop diagram to the renormalization factor connecting renormalized and bare vertex functions in eq. (15) in the creation/annihilation operator language

$$\left[ Z Z_\phi^{-n_{ext}/2} \right]_{\text{Diagram 1}} = \frac{1}{2} \sum_{l=0}^{\infty} \sum_{k,j=0}^{l} I_1(k, l-k) \frac{j!(l-j)!}{k!(l-k)!} a_j^\dagger a_{l_1+l_2-j}^\dagger a_k a_{l-k}. \qquad (77)$$

The factorials follow from the normalization chosen in eq. (6). The contribution in order $\tilde{g}$ gives the operator $X$ in the definition of eq. (17)

$$\begin{aligned}X &= -\frac{1}{6} \sum_{l=0}^{\infty} \sum_{k,j=0}^{l} \frac{1}{l+1} a_j^\dagger a_{l-j}^\dagger a_k a_{l-k} \qquad (78) \\ &\quad \times \left( \frac{1}{\epsilon} + \frac{1}{2} \ln \frac{\mu^2}{(p_j + p_{l-j})^2} - \frac{1}{2} + (S_{l+1} - \frac{1}{2} S_k - \frac{1}{2} S_{l-k}) + O(\epsilon) \right)\end{aligned}$$

Here $p_j$ denotes the momentum carried by the elementary field generated by $a_j^\dagger$. Contributions in order $\tilde{g}^2$ come from diagrams 1,3,4,5,7

$$Y = Y_1 + Y_3 + Y_4 + Y_5 + Y_7 \qquad (79)$$

From eq. (76) one obtains

$$\begin{aligned}Y_1 &= -\frac{N+8}{36} \sum_{l=0}^{\infty} \sum_{k,j=0}^{l} \frac{1}{l+1} a_j^\dagger a_{l-j}^\dagger a_k a_{l-k} \\ &\quad \times \left[ \frac{1}{\epsilon^2} + \frac{1}{2\epsilon} \ln \frac{\mu^2}{(p_j + p_{l-j})^2} - \frac{1}{2\epsilon} + \frac{1}{\epsilon}(S_{l+1} - \frac{1}{2} S_k - \frac{1}{2} S_{l-k}) + O(\epsilon^0) \right]. \quad (80)\end{aligned}$$

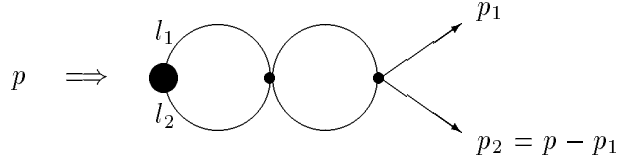

Figure 5: Diagram 3 with an operator insertion with total momentum $p$

## A.2 Diagram 3

The two–loop diagram in fig. 5 has the structure of a product of two one–loop diagrams. The loop integral is therefore simply

$$I_3(l_1, l_2) = I_1(l_1, l_2) \cdot I_1(0, 0). \tag{81}$$

Hence its contribution to the operator $Y$ defined in eq. (79) is

$$\begin{aligned}Y_3 &= \frac{1}{18} \sum_{l=0}^{\infty} \sum_{k,j=0}^{l} \frac{1}{l+1} a_j^\dagger a_{l-j}^\dagger a_k a_{l-k} \\ &\times \left[\frac{1}{\epsilon^2} + \frac{1}{\epsilon} \ln \frac{\mu^2}{(p_j + p_{l-j})^2} + \frac{1}{\epsilon}(S_{l+1} - \frac{1}{2} S_k - \frac{1}{2} S_{l-k}) + O(\epsilon^0)\right]. \end{aligned} \tag{82}$$

## A.3 Diagram 4

With the notation in fig. 6 we have to evaluate the following integral

$$I_4(l_1, l_2) = 32(N+6) \times \left(\frac{g_0^2}{4!^2}\right) \int \frac{d^d q_1}{(2\pi)^d} \frac{d^d q_2}{(2\pi)^d} \frac{(\Delta \cdot (q_1 + p_1))^{l_1} (\Delta \cdot (p_2 - q_1))^{l_2}}{(q_1 + p_1)^2 (q_1 - p_2)^2 q_2^2 (q_1 + q_2)^2}. \tag{83}$$

This calculation is somehow lengthy though standard. Details can be found in ref. [8]. We just quote the result

$$\begin{aligned}I_4(l_1, l_2) &= \tilde{g}^2 \frac{N+6}{18} \frac{l_1! l_2!}{(l_1 + l_2 + 1)!} \\ &\times \left[\left(\frac{1}{2\epsilon^2} + \frac{1}{2\epsilon} \ln \frac{\mu^2}{p^2} + \frac{1}{2\epsilon}(2S_{l_1+l_2+1} - S_{l_1} - S_{l_2}) - \frac{1}{4\epsilon} S_{l_1+l_2+1}\right) [\Delta \cdot p]^{l_1+l_2} \right. \\ &+ \frac{1}{4\epsilon} \sum_{u=1}^{l_1+l_2} \sum_{i,j=0}^{l_1+l_2-u} \frac{1}{u\binom{l_1+l_2}{u}} \binom{j}{u+i-l_2} \binom{l_1+l_2-j}{l_2-i} \binom{l_1+l_2}{j} \\ &\times (\Delta \cdot p_1)^j (\Delta \cdot p_2)^{l_1+l_2-j} \\ &\left. + O(\epsilon^0)\right] + O(\tilde{g}^3). \end{aligned} \tag{84}$$

In creation/annihilation operator language this gives in eq. (79)

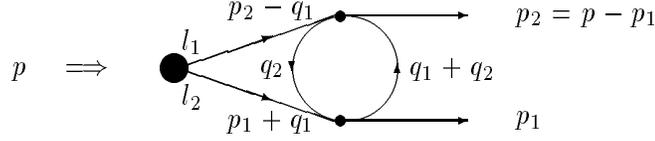

Figure 6: Diagram 4 with an operator insertion with total momentum $p$

$$Y_4 = \frac{N+6}{36} \sum_{l=0}^{\infty} \sum_{k,j=0}^{l} \frac{1}{l+1} a_j^\dagger a_{l-j}^\dagger a_k a_{l-k}$$

$$\times \left[ \frac{1}{2\epsilon^2} + \frac{1}{2\epsilon} \ln \frac{\mu^2}{(p_j + p_{l-j})^2} - \frac{1}{4\epsilon} S_{l+1} + \frac{1}{\epsilon}(S_{l+1} - \frac{1}{2} S_k - \frac{1}{2} S_{l-k}) \right.$$

$$\left. + \frac{1}{4\epsilon} \left( S_l + I(k<j)(S_k - S_j) + I(k>j)(S_{l-k} - S_{l-j}) \right) + O(\epsilon^0) \right] \quad (85)$$

with the notation

$$I(\text{true}) = 1, \quad I(\text{false}) = 0. \quad (86)$$

In deriving eq. (85) from eq. (84) we have used

$$\sum_{u=1}^{l_1+l_2} \sum_{i=0}^{l_1+l_2-u} \frac{1}{u \binom{l_1+l_2}{u}} \binom{j}{u+i-l_2} \binom{l_1+l_2-j}{l_2-i}$$

$$= S_{l_1+l_2} + I(l_1 < j)(S_{l_1} - S_j) + I(l_1 > j)(S_{l_2} - S_{l_1+l_2-j}) \quad (87)$$

as can be shown after some algebra.

### A.4 Diagram 5

With the notation in fig. 7 we have to evaluate the following integral

$$I_5(l_1, l_2, l_3) = 64 \times \left( \frac{g_0^2}{4!^2} \right)$$

$$\times \int \frac{d^d q_1}{(2\pi)^d} \frac{d^d q_2}{(2\pi)^d} \frac{((\Delta \cdot (q_1+q_2))^{l_1}(\Delta \cdot (-q_1))^{l_2}(\Delta \cdot (p-q_2))^{l_3}}{(q_1+q_2)^2 q_1^2 (q_2-p_1)^2 (p-q_2)^2} . \quad (88)$$

Details of the calculation can be found in ref. [8]. $I_5(l_1, l_2, l_3)$ evaluates to

$$I_5(l_1, l_2, l_3) = \frac{\tilde{g}^2}{9} \frac{l_1! l_2! l_3!}{l_1 + l_2 + 1}$$

$$\times \left[ \left( \frac{1}{2\epsilon^2} + \frac{1}{2\epsilon} \ln \frac{\mu^2}{p_2^2} - \frac{1}{2\epsilon} + \frac{1}{2\epsilon} S_{l_1+l_2+1} - \frac{1}{4\epsilon} S_{l_1} - \frac{1}{4\epsilon} S_{l_2} \right) \right.$$

$$\left. \times \sum_{a=0}^{l_1+l_2} \frac{1}{(l_1+l_2-a)!(a+l_3+1)!} (\Delta \cdot p_1)^{l_1+l_2-a} (\Delta \cdot p_2)^{a+l_3} \right.$$

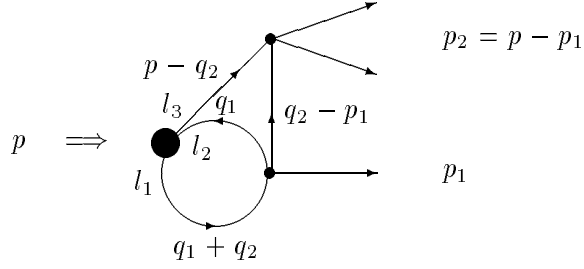

Figure 7: Diagram 5 with an operator insertion with total momentum $p$

$$+\frac{1}{2\epsilon}\sum_{a=0}^{l_1+l_2}\frac{2S_{a+l_3+1}-S_a-S_{l_3}}{(l_1+l_2-a)!(a+l_3+1)!}(\Delta\cdot p_1)^{l_1+l_2-a}(\Delta\cdot p_2)^{a+l_3}$$

$$-\frac{1}{4\epsilon}\sum_{a=0}^{l_1+l_2+l_3}\frac{S_{l_1+l_2+l_3+1}-S_{l_3}}{(l_1+l_2+l_3-a)!(a+1)!}(\Delta\cdot p_1)^{l_1+l_2+l_3-a}(\Delta\cdot p_2)^a$$

$$+\frac{1}{4\epsilon}\sum_{a=0}^{l_3-1}\frac{S_{l_1+l_2+l_3-a}-S_{l_3-a-1}}{(l_1+l_2+l_3-a)!(a+1)!}(\Delta\cdot p_1)^{l_1+l_2+l_3-a}(\Delta\cdot p_2)^a$$

$$+O(\epsilon^0)\bigg] \quad +O(\tilde{g}^3). \tag{89}$$

This gives the contribution $Y_5$ in eq. (79)

$$Y_5=\frac{1}{9}\sum_{l=0}^{\infty}\sum_{k_1,j_1=0}^{l}\sum_{k_2=0}^{l-k_1}\sum_{j_2=0}^{l-j_1}\frac{1}{j_1+j_2+1}\frac{1}{k_1+k_2+1}\,a^\dagger_{j_1}a^\dagger_{j_2}a^\dagger_{l-j_1-j_2}a_{k_1}a_{k_2}a_{l-k_1-k_2}$$

$$\times\bigg[\bigg(\frac{1}{2\epsilon^2}+\frac{1}{2\epsilon}\ln\frac{\mu^2}{(p_{j_1}+p_{j_2})^2}-\frac{1}{2\epsilon}+\frac{1}{2\epsilon}(S_{k_1+k_2+1}-\frac{1}{2}S_{k_1}-\frac{1}{2}S_{k_2})$$

$$+\frac{1}{\epsilon}(S_{j_1+j_2+1}-\frac{1}{2}S_{k_1+k_2+j_1+j_2-l}-\frac{1}{2}S_{l-k_1-k_2})\bigg)I(k_1+k_2+j_1+j_2\geq l)$$

$$-\frac{1}{4\epsilon}(S_{l+1}-S_{l-k_1-k_2})$$

$$+\frac{1}{4\epsilon}(S_{l-j_1-j_2}-S_{l-k_1-k_2-j_1-j_2-1})\,I(k_1+k_2+j_1+j_2<l)\bigg]. \tag{90}$$

## A.5  Diagram 7

Diagram 7 in fig. 8 is trivial since its contribution can be written as a product of one–loop diagrams

$$I_7(l_1,l_2,l_3,l_4)=I_1(l_1,l_2)\cdot I_1(l_3,l_4). \tag{91}$$

This gives the following term in eq. (79)

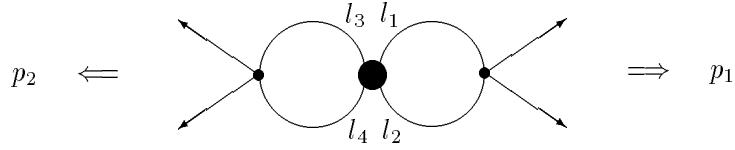

Figure 8: Diagram 7 with an operator insertion with total momentum $p = p_1 + p_2$

$$Y_7 = \frac{1}{72} \sum_{l_1,l_2,l_3,l_4=0}^{\infty} \sum_{j_1=0}^{l_1+l_2} \sum_{j_2=0}^{l_3+l_4} \frac{1}{l_1+l_2+1} \frac{1}{l_3+l_4+1}$$
$$\times a_{j_1}^\dagger a_{l_1+l_2-j_1}^\dagger a_{j_2}^\dagger a_{l_3+l_4-j_2}^\dagger a_{l_1} a_{l_2} a_{l_3} a_{l_4} \qquad (92)$$
$$\times \left( \frac{1}{\epsilon^2} + \frac{1}{\epsilon} \ln \frac{\mu^2}{(p_{j_1}+p_{l_1+l_2-j_1})^2} - \frac{1}{\epsilon} + \frac{1}{\epsilon}(2S_{l_3+l_4+1} - S_{l_3} - S_{l_4}) \right).$$

## Appendix B  Two–point correlation functions

The two diagrams in fig. 9 contribute to the two–point correlation function of two composite operators with the same number of fields $n$ up to order $g$

$$< A_1(r_1)\, A_2(r_2) >_c = \underbrace{< A_1(r_1)\, A_2(r_2) >^{(0)}}_{\text{order } \tilde{g}^0} + \underbrace{< A_1(r_1)\, A_2(r_2) >^{(1)}}_{\text{order } \tilde{g}} + O(\tilde{g}^2). \qquad (93)$$

The operator $A_1$ ($A_2$) has $l_1$ ($l_2$) gradients and a structure as defined in eq. (59). For the renormalized operators follows

$$< A_1^R(r_1)\, A_2^R(r_2) >_c = \left(1 + \frac{\tilde{g}}{\epsilon}\left[\frac{\lambda_{A_1}}{\tilde{g}_c} + \frac{\lambda_{A_2}}{\tilde{g}_c}\right]\right) < A_1(r_1)\, A_2(r_2) >^{(0)}$$
$$+ < A_1(r_1)\, A_2(r_2) >^{(1)} + O(\tilde{g}^2) \qquad (94)$$

with anomalous dimensions $\lambda_{A_1}, \lambda_{A_2}$. It is a standard exercise to evaluate the two diagrams in fig. 9 (for details see ref. [8]). One finds

$$< A_1(r_1)\, A_2(r_2) >^{(0)} = \frac{2^{l_1+l_2+2n(1-\epsilon/2)}}{(4\pi)^{nd/2}} \frac{(\Delta \cdot r)^{l_1}(-\Delta \cdot r)^{l_2}}{(r^2)^{l_1+l_2+n(1-\epsilon/2)}} \times t\, \mathcal{A}^{(0)} \qquad (95)$$

with $r = r_2 - r_1$,

$$t = \sum_{i_1,\ldots,i_n=1}^{N} t^{(1)}_{i_1\ldots i_n} t^{(2)}_{i_1\ldots i_n} \qquad (96)$$

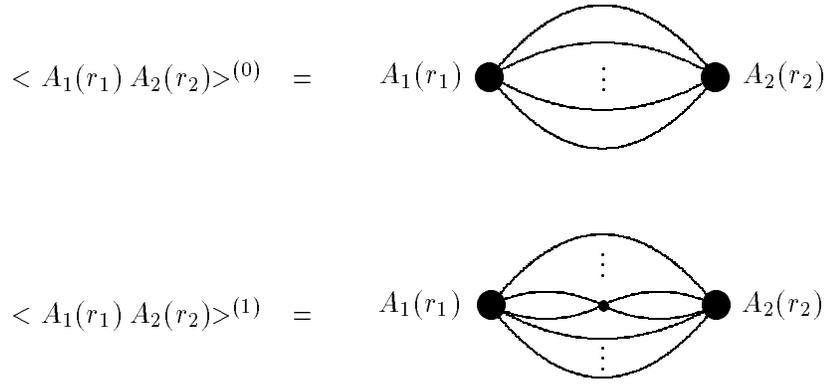

Figure 9: Diagrams contributing to the two–point correlation function of two composite operators up to order $g$.

and the amplitude

$$\mathcal{A}^{(0)} = \sum_{\substack{\{j_1,\ldots,j_n\} \\ \{j'_1,\ldots,j'_n\}}} c^{(1)}_{j_1\ldots j_n} c^{(2)}_{j'_1\ldots j'_n} \sum_{\pi \in S_n} \prod_{s=1}^{n} \frac{\Gamma\left(j_{\pi(s)} + j'_s + 1 - \frac{\epsilon}{2}\right)}{j_{\pi(s)}! j'_s!} \ . \tag{97}$$

Calculating the contribution in order $\tilde{g}$ is more tedious

$$< A_1(r_1)\, A_2(r_2) >^{(1)} = -\frac{4}{3}\tilde{g}\, \frac{2^{l_1+l_2+2n(1-\epsilon/2)}}{(4\pi)^{nd/2}} \frac{(\Delta\cdot r)^{l_1}(-\Delta\cdot r)^{l_2}}{(r^2)^{l_1+l_2+n(1-\epsilon/2)}}$$

$$\times \left(\frac{1}{\epsilon} + \ln(\mu\, r) - \ln 2 - \frac{1}{2}\psi(2) + O(\epsilon)\right) \times t\, \mathcal{A}^{(1)} \tag{98}$$

with

$$\mathcal{A}^{(1)} = \sum_{\substack{\{j_1,\ldots,j_n\} \\ \{j'_1,\ldots,j'_n\}}} c^{(1)}_{j_1\ldots j_n} c^{(2)}_{j'_1\ldots j'_n} \sum_{\substack{i_1 < i_2 \\ i'_1 < i'_2}} \Gamma\left(j_{i_1} + j_{i_2} + j'_{i'_1} + j'_{i'_2} + 2 - \frac{3}{2}\epsilon\right)$$

$$\times \frac{B\left(j_{i_1} + 1 - \frac{\epsilon}{2}, j_{i_2} + 1 - \frac{\epsilon}{2}\right)}{j_{i_1}! j_{i_2}!} \frac{B\left(j'_{i'_1} + 1 - \frac{\epsilon}{2}, j'_{i'_2} + 1 - \frac{\epsilon}{2}\right)}{j'_{i'_1}! j'_{i'_2}!}$$

$$\times \sum_{\pi \in S_{M' \to M}} \prod_{\substack{q=1 \\ q \neq i'_1, i'_2}}^{n} \frac{\Gamma\left(j_{\pi(q)} + j'_q + 1 - \frac{\epsilon}{2}\right)}{j_{\pi(q)}! j'_q!} \ . \tag{99}$$

$S_n$ is the usual permutation group and $S_{M' \to M}$ denotes all mappings between the sets

$$M = \{1,\ldots,n\} \setminus \{i_1, i_2\} \quad , \qquad M' = \{1,\ldots,n\} \setminus \{i'_1, i'_2\}. \tag{100}$$

The $1/\epsilon$–pole terms cancel for one–loop eigenoperators. If one expands the amplitudes in powers of $\epsilon$

$$\mathcal{A}^{(0)} = \mathcal{A}^{(0)}_0 + \epsilon\, \mathcal{A}^{(0)}_\epsilon + O(\epsilon^2)$$
$$\mathcal{A}^{(1)} = \mathcal{A}^{(1)}_0 + \epsilon\, \mathcal{A}^{(1)}_\epsilon + O(\epsilon^2), \tag{101}$$

this gives the relation

$$\left[\frac{\lambda_{A_1}}{\tilde{g}_c} + \frac{\lambda_{A_2}}{\tilde{g}_c}\right] \mathcal{A}_0^{(0)} = \frac{4}{3} \mathcal{A}_0^{(1)}. \tag{102}$$

Combining everything yields for the two–point correlation function of renormalized eigenoperators

$$< A_1^R(r_1) A_2^R(r_2)>_c = \frac{2^{l_1+l_2+2n(1-\epsilon/2)}}{(4\pi)^{n\,d/2}} \frac{(\Delta \cdot r)^{l_1}(-\Delta \cdot r)^{l_2}}{(r^2)^{l_1+l_2+n(1-\epsilon/2)}}$$
$$\times \left\{ \left(1 - \tilde{g}\left[\frac{\lambda_{A_1}}{\tilde{g}_c} + \frac{\lambda_{A_2}}{\tilde{g}_c}\right]\left(\ln(\mu\,r) - \ln 2 - \frac{1}{2}\psi(2)\right)\right) \mathcal{A}_0^{(0)} \right.$$
$$\left. + \left(\epsilon + \tilde{g}\left[\frac{\lambda_{A_1}}{\tilde{g}_c} + \frac{\lambda_{A_2}}{\tilde{g}_c}\right]\right) \mathcal{A}_\epsilon^{(0)} - \frac{4}{3}\tilde{g}\,\mathcal{A}_\epsilon^{(1)} \right\}$$
$$+ \tilde{g}\,O(\epsilon) + O(\tilde{g}^2). \tag{103}$$

[22] K.G. Wilson, Phys. Rev. D7 (1973) 2911

[23] J. Zinn-Justin, Quantum Field Theory and Critical Phenomena (Clarendon Press, Oxford, 1990)
32